\documentclass[pra,10pt,showpacs]{revtex4}
\usepackage{graphicx,amsmath,amssymb}

\newcommand{\beq}{\begin{equation}}
\newcommand{\eeq}{\end{equation}}
\newcommand{\beqa}{\begin{eqnarray}}
\newcommand{\eeqa}{\end{eqnarray}}

\newcounter{sub}
\newcounter{subeqn}[sub]
 \stepcounter{sub}

\begin{document}

\def\zzeta{\zeta\hspace{-0.21cm}\zeta}
\def\eeta{\eta\hspace{-0.16cm}\eta}


\title{Phase Space Quantum Mechanics - Direct} \centerline{ \Huge  }
\author{S. Nasiri  $^{1,2}$ }
\author{Y. Sobouti $^{1}$}
\author{ F. Taati  $^{1,3}$}
\affiliation{ $^1$ Institute for Advanced Studies in Basic
Sciences,
Zanjan, 45195-1159, Iran \\
 $^2$Department of Physics, Zanjan University Zanjan
Iran \\$^3$Department of Physics, University of Kurdistan, D-78457
Sanadaj, Iran
\\\footnote{htaati@iasbs.ac.ir\\nasiri@iasbs.ac.ir\\ sobouti@iasbs.ac.ir
}}

\date{\today}

\begin{abstract}
Conventional approach to quantum mechanics in phase space,
$(q,p)$, is to take the operator based quantum mechanics of
Schr\"{o}dinger, or and equivalent, and assign a $c$-number
function in phase space to it. We propose to begin with a higher
level of abstraction, in which the independence and the symmetric
role of $q$ and $p$ is maintained throughout, and at once arrive
at phase space state functions. Upon reduction to the $q$- or
$p$-space the proposed formalism gives the conventional quantum
mechanics, however, with a definite rule for ordering of factors
of non commuting observables. Further conceptual and practical
merits of the formalism are demonstrated throughout the text.
\end{abstract}


\maketitle

\section{Introduction}
\label{intro} Wigner's 1932 initiative \cite{E. Wigner} is a
reformulation of the operator based quantum theory of
Schr\"{o}dinger in the language of c-number distribution functions
in a phase space. His prescription, however, turns out to have a
feature extra to what one finds in Schr\"{o}dinger's theory. There
is nothing in the founding principles of the operator based theory
to prescribe a rule for ordering of the factors of non-commuting
operators in a product. In contrast, Wigner's formalism, upon
reduction from phase space to the configuration space, acquires
Weyl's ordering \cite{H.Weyl, M. Hillery and R. F. O'Connell and
M. O. Scully and E. Wigner}. How and at what stage, in going from
Schr\"{o}dinger's state functions in configuration space to those
of Wigner in phase space and again coming back to the
configuration space, Weyl's ordering creeps in? This feature is
not unique to Wigner's functions. Other distributions exist in the
literature, e.g., Kirkwood  \cite{J.G. Kirkwood}, Husimi \cite{K.
Husimi}, Margenau and Hill \cite{H. Margenau and R.N. Hill}, etc.
Each of them carries its own ordering rule, with no precedence in
the configuration space formalism. Can one conjecture that the
phase space formulations of quantum mechanics are more complete
than their configuration space counterpart, because of their
built-in ordering rules? If so, there should be a way to arrive at
phase space formulations without reference to the conventional
operator based theory. Here we argue that in the classical
dynamics and classical statistical dynamics (Liouville's equation)
the generalized coordinates and momenta, $q$ and $p$,
respectively, play symmetric and more importantly, independent
roles. In the operator based quantum theory one or the other
looses its identity at the expense of the other and the formalism
reduces, to one in either $q$ or $p $ space. One could avoid this
by carrying the $q$ and $p$ formalisms concomitantly and at once
arrive at state functions in $~qp~$spaces. The so-obtained state
functions are the $~qp~$representation of the mixed states of
quantum statistical mechanics. The operator based theory emerges
as a special case of this general one, but this time with a
definite ordering rule for non-commutative operators. The rule
depends on the nature of the $q$ and $p$ variables, adopted
initially.
\section{Extension of the Classical Dynamics}
Let $~q=\{q_i(t), i=1,..,N\}~$be the collection of the generalized
coordinates describing the state of motion of a dynamical system.
It is customary to assign a lagrangian, $L^{q}(q,\dot{q})$, to the
system, define the conjugate momenta, $p=\partial L^{q}/\partial
\dot{q}$, and construct the $~H(q,p)=\dot{q}p-L^{q}$. One may do
the other way around. Begin with a given$~H(q,p)~$and
find$~L^{q}(q,\dot{q})~$as a solution of the following
differential equation,
\begin{equation}
H(q,\frac{\partial L^{q}}{\partial \dot{q}})-\dot{q}\frac{\partial
L^{q}}{\partial \dot{q}}+L^{q}=0.\label{eq1}
\end{equation}
\stepcounter{sub}
 One may, however, carry out the same procedure
with $q$ replaced by $p$ and arrive at a $L^{p}(p,\dot{p})$
satisfying the differential equation
\begin{equation}
H(\frac{\partial L^{p}}{\partial \dot{p}},p)+\dot{p}\frac{\partial
L^{p}}{\partial \dot{p}}-L^{p}=0.\label{eq2}
\end{equation}
\stepcounter{sub}
 The use of$~L^{p}~$to study the evolution of a
dynamical system is not a common practice. But it is a possibility
and has precedence \cite{GPS}. There is no barring to employ the
two alternatives simultaneously. We follow Sobouti and Nasiri
\cite{Y. Sobouti and S. Nasiri}, (hereafter, paper I) and define
the "extended lagrangian"
\begin{eqnarray}
{\cal L}(q,\dot{q};
p,\dot{p})=-\dot{q}p-q\dot{p}+L^{q}(q,\dot{q})+L^{p}(p,\dot{p}).\label{eq3}
\end{eqnarray}
\stepcounter{sub}
 The first two terms on the right hand side
constitute a total time derivative and are introduced for later
convenience. One may now write down the Euler-Lagrange equations
for $q$ and $p$,
 \setcounter{subeqn}{1}
\begin{eqnarray}
&& \frac{d}{dt}\frac{\partial {\cal L}}{\partial
\dot{q}}-\frac{\partial {\cal L}}{\partial
q}=\frac{d}{dt}\frac{\partial  L^{q}}{\partial
\dot{q}}-\frac{\partial L^{q}}{\partial q}=0 ,\label{eq4}\\
\stepcounter{subeqn}
 &&\frac{d}{dt}\frac{\partial {\cal
L}}{\partial \dot{p}}-\frac{\partial {\cal L}}{\partial
p}=\frac{d}{dt}\frac{\partial  L^{p}}{\partial
\dot{p}}-\frac{\partial L^{p}}{\partial p}=0.\label{eq5}
\end{eqnarray}
\stepcounter{sub}
 Equation (\ref{eq4}) is the conventional
equation of motion in $q$ space. With preassigned initial values
$q(t_0)$ and $\dot{q}(t_0)$ at $t_0$ it can be solved for the
orbits $q(t)$ in $q$ space. Similarly, with given initial values
$p(t_0)$and $\dot{p}(t_0)$, Eq (\ref{eq5})  can be solved for the
orbits $p(t)$ in the $p$ space. The condition for $q$ and $p$
orbits to represent the same state of motion of the system are
$p(t_0)=\partial L^q/\partial \dot{q}|_{t_0}$ and $q(t_0)=\partial
L^p/\partial \dot{p}|_{t_0}$. Such a state of motion will be
referred to as a "pure state". Otherwise it will be called a
"mixed state" of motion. The nomenclature is from the statistical
quantum mechanics and it will be seen later that they imply the
same notions as therein. On a pure state $p$ and $q$ are initially
canonically conjugate pairs and it is shown in paper I that once
they are canonically conjugate at one time they remain so for all
times. On the other hand there are no restrictions on the initial
values of $q$ and $p$ on mixed states. Therefore, $q(t)$ and
$p(t)$ remain unrelated and evolve independently. The existence of
the extended lagrangian ${\cal L}(q,\dot{q};p,\dot{p})$, however,
permits the following "extended momenta" to be defined,
\setcounter{subeqn}{1}
\begin{eqnarray}
&&\pi_q=\frac{\partial {\cal L}}{\partial \dot{q}}=\frac{\partial
L^q}{\partial \dot{q}}-p, \label{eq6}\\
 \stepcounter{subeqn}
&&
\pi_p=\frac{\partial {\cal L}}{\partial \dot{p}}=\frac{\partial
L^p}{\partial \dot{p}}-q. \label{eq7}
\end{eqnarray}
\stepcounter{sub}
 These in turn allow an "extended hamiltonian" to
be defined through the following Legendre transformation,
\begin{equation}
{\cal H}(q,\pi_q;p,\pi_p)=\dot{q}\pi_q+\dot{p}\pi_p-{\cal
L}(q,\dot{q};p,\dot{p}). \label{eq8}
\end{equation}
\stepcounter{sub}
 To eliminate $\dot{q}$ and $\dot{p}$ from ${\cal
H}$ one substitutes a) for ${\cal L}$ from Eq (\ref{eq3}), b) for
$L^q$ and $L^p$ from Eqs (\ref{eq1}) and (\ref{eq2}) and  c) for
$\partial L^q/\partial \dot{q}$ and $\partial L^p/\partial
\dot{p}$ from Eqs (\ref{eq6}) and (\ref{eq7}). One arrives at
\begin{eqnarray}
{\cal H}(q, \pi_q ;p, \pi_p)=H(q ,p+\pi_q)-H(q+\pi_p~, p)\nonumber\\
=\sum_{n=0}\frac{1}{n!}\left[\frac{\partial^nH}{\partial
p^n}\pi_q^n-\frac{\partial^nH}{\partial q^n}\pi_p^n
\right],\label{eq9}
\end{eqnarray}
\stepcounter{sub}
 where the derivatives are to be evaluated at
$(q,p)$. We leave it to the reader to familiarize him/herself with
${\cal H}$ by writing down four Hamilton's equations for
$\dot{q},~\dot{\pi_q},~\dot{p}$ and $\dot{\pi_p}$~. Here, the
condition for pure state motions is $\pi_{q}(t_0)=\pi_{p}(t_0)=0$,
and once they are initially zero they remain so for all times.
Then by Eqs (\ref{eq6}) and (\ref{eq7}) $q$ and $p$ turn into
canonically conjugate pairs for all times (paper I). To summarize,
for any dynamical system we introduce an extended phase space,
$(q,~\pi_q;~p,~\pi_p)$, extended momenta, lagrangians, and
hamiltonians. All concepts and procedures of the conventional
dynamics are extendible to this extended dynamics. Of particular
relevance to this paper, that will be referred to shortly, are: 1)
canonical transformations from one set of variables
$(q,~\pi_q;~p,~\pi_p)$ to another, and 2) Poisson's brackets
extended as
\begin{eqnarray}
\{F, G\}=\frac{\partial F}{\partial q}\frac{\partial G}{\partial
\pi_q}-\frac{\partial F}{\partial \pi_q}\frac{\partial G}{\partial
q}+\frac{\partial F}{\partial p}\frac{\partial G}{\partial
\pi_p}-\frac{\partial F}{\partial \pi_p}\frac{\partial G}{\partial
p}.\label{eq10}
\end{eqnarray}
\stepcounter{sub}
\section{Quantum Dynamics in  $\textbf{qp}$ space}
Now that we have the extended hamiltonian of Eq (\ref{eq9}) we may
construct a quantum mechanics in $qp$ space. We do this on the
following premises:
\begin{enumerate}
    \item
    Let ${\cal
X}$ be the function space of all integrable complex
 functions $\chi(q,p)$. Let $q,~\pi_q,~p$, and, $\pi_p$ be operators on ${\cal
 X}$, satisfying the commutation rules
\begin{equation}
[q, \pi_q]=[p, \pi_p]=i\hbar,~ [q,p]=[\pi_q,\pi_p]=[q, \pi_p]=[p,
\pi_q]=0.\label{eq11}
\end{equation}
\stepcounter{sub}
 These are the fundamental Poisson brackets of Eq
(\ref{eq10}), promoted to commutation brackets by Dirac's
prescription. Note the manifest independence of $q$ and $p$ in the
vanishing of their commutation brackets.
    \item
    By the virtue of Eq (\ref{eq11}), ${\cal H}$ is now an operator
on ${\cal X}$. Let $\chi(q,p,t)\in{\cal X}$ be a state function
satisfying the Schr\"{o}dinger-like equation
\begin{equation}
i\hbar\frac{\partial \chi}{\partial t}={\cal
H}\chi=\left[H(q,p-i\hbar \frac{\partial}{\partial q})-H(q-i\hbar
\frac{\partial}{\partial p},p)\right]\chi.\label{eq12}
\end{equation}
\stepcounter{sub}
    \item
Let the rule to evaluate the expectation values of an observable
$O(q,p)$, a real c-number operator on ${\cal X}$, be
\begin{equation}
<O(q,p)>=\int O(q,p)~{\rm Re}~\chi dqdp=\frac{1}{2}\int
O(q,p)(\chi+\chi^*)dqdp.\label{eq13}
\end{equation}
\stepcounter{sub}
 We will return to this averaging rule shortly,
and revise it. The logic behind it, however, is to be noted, the
averages of observables should be real. In what follows we
demonstrate that, 1) the formalism so designed is a theory of
quantum ensembles in phase space. Its pure state case is the
conventional quantum mechanics, however, with a definite ordering
rule accompanying it. 2) It can be transformed to other phase
space formalisms, including to that of Wigner, by suitable unitary
or similarity transformations on ${\cal X}$. The latter in turn
originates from suitable canonical transformations from one
extended phase space coordinates to another.
\end{enumerate}
 \subsection{Solutions of Equation (\ref{eq12})}
To begin with, $\chi$ is of the form
\begin{equation}
\chi(q,p)=F(q,p)e^{-ipq/\hbar}.\label{eq14}
\end{equation}
\stepcounter{sub}
 The exponential factor is a consequence of the
total time derivative, $-d(qp)/dt$ in Eq (\ref{eq3}). It is easily
verified that
 \setcounter{subeqn}{1}
\begin{eqnarray}
&&(p-i\hbar \frac{\partial}{\partial q})\chi=i\hbar \frac{\partial
F}{\partial q}e^{-ipq/\hbar},\label{eq15}\\
 \stepcounter{subeqn}
&&(q-i\hbar \frac{\partial}{\partial p})\chi=i\hbar \frac{\partial
F}{\partial p}e^{-ipq/\hbar}.\label{eq16}
\end{eqnarray}
\stepcounter{sub}
 Substitution of Eqs (13)
and (\ref{eq14}) in Eq (\ref{eq12}) gives
\begin{equation}
i\hbar\frac{\partial F}{\partial t}=\left[H(q,-i\hbar
\frac{\partial}{\partial q})-H(-i\hbar\frac{\partial}{\partial
p},p)\right]F.\label{eq17}
\end{equation}
\stepcounter{sub}
 The operators on the right hand side of Eq
(\ref{eq17}) are recognized as the hamiltonians of the
conventional quantum mechanics, the first in $q$ and the second in
$p$ representation. Thus, one obtains the superposition of the
separable solutions
\begin{equation}
\chi(q,p,t)=\sum_{\alpha,\beta}A_{\alpha\beta}\psi_\alpha(q,t)\phi_\beta^*(p,t)e^{-ipq/\hbar},\label{eq18}
\end{equation}
\stepcounter{sub}
 where
 \setcounter{subeqn}{1}
\begin{eqnarray}
&&i\hbar\frac{\partial \psi_\alpha}{\partial t}=H(q,-i\hbar~
\frac{\partial}{\partial q})\psi_\alpha,\label{eq19}\\
\stepcounter{subeqn} &&i\hbar\frac{\partial \phi_\beta}{\partial
t}=H(i\hbar~ \frac{\partial}{\partial
p},~p)\phi_\beta.\label{eq20}
\end{eqnarray}
\stepcounter{sub}
 To each $\psi_\alpha(q)$ there corresponds a
$\phi_\alpha(p)$ that are Fourier transforms of each other,
\begin{equation}
\psi_\alpha(q)=\frac{1}{(2\pi \hbar)^{N/2}}\int
\phi_\alpha(p)e^{ipq/\hbar}dp~,\label{eq21}
\end{equation}
\stepcounter{sub}
 where N is the number of degrees of freedom of
the system.
\subsection{The averaging rule revisited: Acceptable state functions}
 Let $Q(q)$ be an observable represented by a real polynomial or series
in $q$. Its matrix representation, $\hat{Q}$, in either $\chi$-,
$\psi$-, or $\phi$-basis is hermitian. Thus
\begin{eqnarray}
Q_{\beta\alpha}&=&\int\psi_\alpha(q)Q(q)\phi^*_\beta(p)e^{-ipq/\hbar}dp~dq\nonumber\\&=&
\int\psi^*_\beta(q)Q(q)\psi_\alpha(q)dq \nonumber\\&=&\int
\phi_{\beta}^*(p)Q(i\hbar \frac{\partial}{\partial
p})\phi_\alpha(p)dp=Q^*_{\alpha \beta},\label{eq22}
\end{eqnarray}
\stepcounter{sub}
 where we have used the fact that $\psi$ and
$\phi$ bases are the Fourier transforms of each other. The
coefficient $(2\pi \hbar)^{-N/2}$ is suppressed for brevity. The
expectation value of $Q$, by Eq (\ref{eq13}), now becomes
\begin{equation}
<Q>=\frac{1}{2}\int Q(\chi+\chi^*)dp
dq=\frac{1}{2}\rm{tr}[\hat{Q}(\hat{A}+\hat{A^\dag})],\label{eq23}
\end{equation}
\stepcounter{sub}
 where $\hat{A}$ is the matrix of
$A_{\alpha\beta}$ of Eq (\ref{eq18}). This gives the freedom of
choosing $\hat{A}=\hat{A^\dag}$ and of simplifying Eq (\ref{eq13})
to read $<Q>=\int Q\chi dp~ dq=$tr$(\hat{Q}\hat{A})$. Choosing
$Q(q)=1$, imposes the further restriction tr$\hat{A}=1$. Requiring
the averages of all positive definite functions of $q$ to be
positive still restricts $\hat{A}$ to be a positive definite
matrix. Had one chosen a differentiable function $P(p)$ instead of
$Q(q)$, one still would have arrived at the same requirements for
$\hat{A}$. To summarize, $\chi$ of Eq (\ref{eq18}) is a physically
acceptable solution if
\begin{equation}
\hat{A}=\hat{A^\dag},~~\rm{positive~~
definite,~and}~\rm{tr}\hat{A}=1.\label{eq24}
\end{equation}
\stepcounter{sub}
 With this provision the averaging rule of Eq
(\ref{eq13}) for $Q(q)+P(p)$ reduces to
\begin{equation}
<Q(q)+P(p)>=\int(Q+P)\chi dp~ dq.\label{eq25}
\end{equation}
\stepcounter{sub}
For a product $Q(q)P(p)$, by the prescription of Eq (\ref{eq13})
and with the restrictions of Eq (\ref{eq24}) on $\hat{A}$, one has

\begin{equation}
<QP>=\rm{Re
~tr}(\hat{Q}\hat{P}\hat{A})=\frac{1}{2}~\rm{tr}(\hat{Q}\hat{P}\hat{A}+\hat{A}\hat{P}\hat{Q})
=\rm{tr}[\frac{1}{2}(\hat{Q}\hat{P}+\hat{P}\hat{Q})\hat{A}],\label{eq26}
\end{equation}
\stepcounter{sub}
 where $\hat{Q}$ and $\hat{P}$ are the matrix
representations of $Q(q)$ and $P(p)$ as in Eq (\ref{eq22}).
Translation of this to the $q$ space language, say, is
\begin{equation}
<QP>=\frac{1}{2}~A_{\alpha\beta}\int \psi_\beta^*(q)[Q(q)P(-i\hbar
\frac{\partial}{\partial q})+P(-i\hbar \frac{\partial}{\partial
q})Q(q)]\psi_\alpha dq.\label{eq27}
\end{equation}
\stepcounter{sub}
 Thus, upon reduction of the formalism of the
present paper to that of the $q$-space, the ordering rule
associated with a product $Q(q)P(p)$ is the symmetric ordering. It
has emerged from the formalism itself, unlike the ad hoc ordering
rules of the conventional quantum mechanics.
\section{More about Equation (\ref{eq12})}
It was stated earlier that the proposed dynamics is essentially
 that of the ensembles. Here we elaborate on this, and
 show that
 1)the classical limit of the theory is Liouville's
 equation that governs the dynamics of classical ensembles. 2) Its
 pure state case is Schr\"{o}dinger's operator based theory. 3) In
 its full generality the theory gives von Neumann's density matrix and the
 evolution equation associated with it.
\subsection{Classical correspondence}
In Eq (\ref{eq12}) expanding the hamiltonian operators about
$(q,p)$, and retaining only the first terms in the expansion,
gives
\begin{equation}
\frac{\partial \chi}{\partial t}+\frac{\partial H}{\partial
p}\frac{\partial \chi}{\partial q}-\frac{\partial H}{\partial
q}\frac{\partial \chi}{\partial p}=\frac{d\chi}{dt}=0.\label{eq28}
\end{equation}
\stepcounter{sub}
 This is the Liouville equation for the
distribution function of classical ensembles. Its most general
solutions are $\chi[q(t),p(t)]$, where $q(t)$ and $p(t)$ are the
classical trajectories in $q$ and $p$ spaces. The two trajectories
may represent the same state of motion if they satisfy the
conditions of initial canonical conjugacy  narrated below Eqs (4)
. Otherwise, they remain independent and evolve independently. It
is this classical notion of independence that we have carried
through to the quantum formalism. Let us also note that the
reduction of the phase space evolution equation to the classical
Liouville's equation is a common feature of all such formalisms.
\subsection{Schr\"{o}dinger's case}
Allowance for only one term in Eq (\ref{eq18}) reproduces the
conventional quantum mechanics in all its details. Thus
\setcounter{subeqn}{1}
\begin{eqnarray}
&&\chi=\psi(q)\phi^*(p)e^{-ipq/\hbar},\label{eq29}\\
\stepcounter{subeqn} &&
 i\hbar\frac{\partial\psi}{\partial t}=H(q,
-i\hbar\frac{\partial}{\partial
 q})\psi,\label{eq30}\\
 \stepcounter{subeqn}
&&\rm{\psi~~and ~~\phi~~Foureir ~~transforms ~~of~~ each
 ~~other},\label{eq31}\\
 \stepcounter{subeqn}
 &&\int\chi dp~dq=\int\psi^*\psi dq=\int\phi^*\phi
 dp=1,\label{eq32}\\
 \stepcounter{subeqn}
 &&<Q(q)P(p)>=\frac{1}{2}\int\psi^*[P(-i\hbar\frac{\partial}{\partial q})Q(q)
 +Q(q)P(-i\hbar\frac{\partial}{\partial q})]\psi dq.\label{eq33}
\end{eqnarray}
\stepcounter{sub}
 Heisenberg's uncertainty principle follows immediately from Eqs
(25) that one may find in standard texts in quantum mechanics. The
ordering rule of Eq (\ref{eq33}) is, however, the added feature of
the theory.
\subsection{Density matrix and Von Neumann's equation}
The state function of Eq (\ref{eq18}), as it stands represents the
 state of an ensemble in a mixed state. If the matrix $\hat{A}$ is
 diagonalized to $A_{\alpha\beta}=A_\alpha\delta_{\alpha\beta}$,
 $\chi$ reduces to $\chi=\sum A_\alpha\psi_\alpha\phi^*_\alpha
 e^{-ipq/\hbar}$. Upon integration over $q$ or $p$ one immediately
 recognizes $A_\alpha $ as the probability of the system to be in
 the state $\psi_\alpha(q,t)$ or $\phi_\alpha(p,t)$. One may
 however do better. Let $\{\psi_n(q)\}$ be a complete orthonormal
time independent basis set, and $\{\phi_n(p)\}$ be its Fourier
replica. These basis sets are not required to be the eigenstates
of $H(q,p)$, though this is a possibility. Hereafter, to avoid the
ambiguity, we use the latin subscripts to denote the members of
the basis set and reserve greek subscripts to denote the solutions
of Eqs (\ref{eq19}) and (\ref{eq20}). Expansion of $\chi$ in these
bases assumes the form
$\chi(q,p,t)=A_{mn}(t)\psi_n(q)\phi_m^*(p)e^{-ipq/\hbar}$.
Substituting this form in Eq (\ref{eq12}), multiplying the
resulting equation by $\psi^*_n(q)\phi_m(p)e^{ipq/\hbar}$, and
integrating over $q$ and $p$ gives
\begin{equation}
i\hbar\frac{d\hat{A}}{dt}=[\hat{A},\hat{H}],~~~~\hat{A}=\hat{A}^\dag
~~\rm{positive~ definite~and~}\rm{tr}\hat{A}=1,\label{eq34}
\end{equation}
\stepcounter{sub}
 where $\hat{A}$ is the matrix of the expansion
coefficients and $\hat{H}$ that of the  $H(q,p)$ in either
$\chi$-, $\psi$- or $\phi$- basis. Equation (\ref{eq34}) is von
Neumann's equation for the evolution of the density matrix. As is
known the case tr$(\hat{A}^2)$=tr$\hat{A}=1$ represents an
ensemble in a pure state. If tr$(\hat{A}^2)<1$, the ensemble is in
a mixed state.
\section{Canonical transformations}
All machinery of the canonical transformations from one extended
coordinate system to another and their associated unitary or
similarity transformations in the function space are available for
a forage of deliberations. Except for a passing remark on the
prospects of fuller uses of this approach at the end of this
section, here we confine ourselves to one one-parameter family of
transformations of which Wigner's state function emerges as a
special case. Husimi's all positive distribution functions are
also briefly mentioned.

 Consider the infinitesimal transformations
\begin{equation}
q=Q-\delta\alpha\Pi_P,~\pi_q=\Pi_Q;~p=P-\delta\alpha\Pi_Q
,~\pi_p=\Pi_P,\label{eq35}
\end{equation}
\stepcounter{sub}
 The generator of the transformation is $G=\pi_p\pi_q$. To this (and for a finite $\alpha$) there
corresponds the unitary operator
\begin{equation}
U_{\alpha}=e^{\frac{-i\alpha
G}{\hbar}}=e^{i\hbar\alpha\frac{\partial^2}{\partial q
\partial p}},~~ U_{\alpha}^\dag U_{\alpha}=1,\label{eq36}
\end{equation}
\stepcounter{sub}
 in the function space. Operating by $U_{\alpha}$ on a pure state
function $\chi(q,p,t)=\psi(q)\phi^*(p)\exp(-ipq/\hbar)$ generates
another state function (let us call it $\alpha$-representation)
\begin{equation}
\chi_\alpha(q,p,t)=U_{\alpha}\chi=(\frac{1}{2\pi\hbar})^N\int
\psi(q-\alpha\tau)\psi^*(q+(1-\alpha)\tau)e^{ip\tau/\hbar}d\tau.\label{eq37}
\end{equation}
\stepcounter{sub}
 See Appendix for proof of Eq (\ref{eq37}). For
$\alpha=1/2$, Eq (\ref{eq37}) gives Wigner's standard function
\cite{R. F. O'Connell and L. Wang, J. E. Moyal},
$\chi_{1/2}=W(q,p,t)$. The cases $\alpha=0$ and $1$ simply give
back $\chi$ and $\chi^*$ of this paper, respectively. Similarly,
operation by $U_{\alpha}$ on Eq (\ref{eq12}) gives the evolution
equation for $\chi_{\alpha}$
\begin{eqnarray}
i\hbar\frac{\partial \chi_{\alpha}(q,p,t)}{\partial
t}&=&i\hbar\frac{\partial}{\partial
t}(U_\alpha\chi)=(U_\alpha{\cal H}U_\alpha^\dag)U_\alpha
\chi\nonumber,\\i\hbar\frac{\partial
\chi_{\alpha}(q,p,t)}{\partial t}&=&{\cal
H}_\alpha\chi_\alpha=-\frac{{\hbar}^2(1-2\alpha)}{2m}\frac{\partial^2}{\partial
q^2}\chi_{\alpha}-i\hbar\frac{p}{m}\frac{\partial}{\partial
q}\chi_{\alpha}+\sum_{n=0}\frac{(-\alpha)^n-(1-\alpha)^n}{n!}(-i\hbar)^n
\frac{\partial^{n}V}{\partial q^{n}}\frac{\partial^n}{\partial
p^n}\chi_{\alpha}.\label{eq38}
\end{eqnarray}
\stepcounter{sub}
 See Appendix for proof of Eq (\ref{eq38}). For $\alpha=1/2$,
even $n$ terms in Eq (\ref{eq38}) cancel out and one again
recovers Wigner's evolution equation, \cite{M. Hillery and R. F.
O'Connell and M. O. Scully and E. Wigner}. See Eq (A7).
 \subsection{Assigning $q$-space operators to phase space functions; ordering rule}
The phase space state functions are devised to evaluate the
expectation values of a c-number observable, $F(q,p)$, by
integrations over the phase space. Upon reduction to the $q$
space, say, $f(q,p)$ turns into  a differential operator in terms
of $q$ and $\pi_q$. The questions are: 1)how different factors of
non commuting $q$ and $\pi_q$ are ordered in a given
$\alpha$-representation? 2)Averaging a given $F(q,p)$ with
different $\chi_\alpha$'s gives different values, how such
averages change from one $\alpha$-representation to another? Let
$\hat{F}_\alpha(q,\pi_q)$ be the $q$ space operator corresponding
to the $c$-number monomial $q^np^m$ in phase space when averaged
by $\chi_\alpha$. The defining equation for
$\hat{F}_\alpha(q,\pi_q)$ is
\begin{equation}
<q^n p^m>_\alpha=\int q^np^m\chi_\alpha dp
dq=\int\psi^*(q)\hat{F}_\alpha(q,\pi_q)\psi(q)dq.\label{eq39}
\end{equation}
\stepcounter{sub}
 For the combination of $\alpha=0$ and 1
corresponding to ($\chi +\chi^*$) of Eq (\ref{eq13}) this is
already worked out in Eq (\ref{eq33}) and is the symmetric
ordering
\begin{equation}
q^np^m\rightarrow \frac{1}{2}(q^n\pi_q^m+\pi_q^mq^n).\label{eq40}
\end{equation}
\stepcounter{sub}
 For a general $\alpha$, it is given in Appendix
Eq (A10)
\begin{equation}
q^np^m\rightarrow \sum_{r=0}^m\left(%
\begin{array}{c}
  m \\
  r \\
\end{array}%
\right) ((1-\alpha)\pi_q)^{r}q^n(\alpha\pi_q)^{m-r}.\label{eq41}
\end{equation}
\stepcounter{sub}
 For $\alpha=1/2$ this reduces to Weyl's ordering
\cite{H.Weyl, M. Hillery and R. F. O'Connell and M. O. Scully and
E. Wigner} which is known to go with Wigner's functions.
 To answer the second question we note the following
\begin{eqnarray}
<q^np^m>_{\alpha=0}=\int q^np^m\chi dq~dp=\int
q^np^mU^\dag_\alpha\chi_\alpha dq~dp=\int U_\alpha(
q^np^m)\chi_\alpha dq~dp=<U_\alpha(q^np^m)>_\alpha,\label{eq42}
\end{eqnarray}
\stepcounter{sub}
 where by Eq (\ref{eq37}) we have used
$\chi=U^\dag\chi_\alpha$. The conclusion is that $q^np^m$ averaged
by $\chi$ is the same as $U_\alpha(q^np^m)$ averaged by
$\chi_\alpha$. Upon adoption of
$U_\alpha=\sum\frac{(-i\alpha\hbar)^k}{k!}
\frac{\partial^k}{\partial q^k}\frac{\partial^k}{\partial p^k}$
and operation by it on $q^np^m$ one finds
\begin{eqnarray}
U_\alpha(q^np^m)&=&\sum_{k=0}^{smaller~ of~ n~ or~
m}(-i\hbar\alpha)^kk!\left(%
\begin{array}{c}
  n \\
  k \\
\end{array}%
\right)\left(%
\begin{array}{c}
  m \\
  k \\
\end{array}%
\right)q^{n-k}p^{m-k}\nonumber\\&
=&q^np^m+(-i\hbar\alpha)mnq^{n-1}p^{m-1}+\frac{1}{2}(-i\hbar\alpha)^2n(n-1)m(m-1)q^{n-2}p^{m-2}+...\label{eq43}
\end{eqnarray}
\stepcounter{sub}
\subsection{Assigning phase space functions to $q$ space operators}
To a given operator $\hat{F}(q,\pi_q)$, a Taylor-expanded series
in whatever order of powers of $q$ and $\pi_q$, we associate the
following c-number function
\begin{equation}
F(q,p)=\sum_{n,m}F_{mn}\chi_{nm}=<q|\hat{F}|p>e^{-\frac{ipq}{\hbar}},\label{eq44}
\end{equation}
\stepcounter{sub}
 where $F_{mn}=<n|\hat{F}|m>$, is the matrix
element of $\hat{F}$ in the basis of the eigenstates of
$\hat{H}(q,\pi_q)$. This is actually the inverse of the procedure
that we used in Eq (\ref{eq39}) to associate an operator with a
c-number function (let $\alpha=0$ and replace $q^np^m$ by $F(q,p)$
in Eq (\ref{eq41}) to see the analogy). The second equality in Eq
(\ref{eq44}) expresses the same in the ket- and bra- notation of
Dirac.

The corresponding function in $\alpha$-representation is simply
\begin{eqnarray}
F_\alpha(q,p)=U_\alpha F(q,p)=\sum_{n,m}F_{mn}U_\alpha\chi_{nm}=
\int
<q-\alpha\tau|\hat{F}|q+(1-\alpha)\tau>e^{ip\tau/\hbar}d\tau.\label{eq45}
\end{eqnarray}
\stepcounter{sub}
 This is actually the generalization of Eq
(\ref{eq37}) for a general operator $\hat{F}(q,\pi_q)$. The rule
for the product $\hat{F}=\hat{A}\hat{B}$ is worked out in Eq
(A13):
\begin{eqnarray}
F(q,p=)<q|\hat{A}(q,\pi_q)\hat{B}(q,\pi_q)|p>e^{-ipq/\hbar}=
\sum_{n=0}^{\infty}\frac{(-i\hbar)^n}{n!}\frac{\partial^nA(q,p)}{\partial
p^n}\frac{\partial^nB(q,p)}{\partial q^n}.\label{eq46}
\end{eqnarray}
\stepcounter{sub}
 One may also work out the
$\alpha$-representation of Eq (\ref{eq46}):
\begin{eqnarray}
F_\alpha(q,p)=U_\alpha F(q,p)&=&A_\alpha\left[ q+i\hbar\alpha
\frac{\partial}{\partial
p},p-i\hbar\alpha(1-\alpha)\frac{\partial}{\partial
q}\right]B_\alpha(q,p)\nonumber\\&=&\sum_{n=0}^{\infty}
\frac{(i\hbar)^n}{n!}\left[\alpha\frac{\partial }{\partial
q_A}\frac{\partial}{\partial p_B}-(1-\alpha)\frac{\partial
}{\partial p_A}\frac{\partial}{\partial
q_B}\right]^nA_\alpha(q,p)B_\alpha(q,p).\label{eq47}
\end{eqnarray}
\stepcounter{sub}
 See Eq (A17) for details of the
derivation. In section VI, we analyze Bloch's problem as an
illustration of the use of the developments of the last two
subsections.

    \subsection{A remark on general transformations:}
    An economical way of treating canonical transformations is the
symplectic formalism. Let $\eeta$ be the column vector
$(q,~\pi_q;~p,~\pi_p)$. The  equations of the classical dynamics
assume the following form
\begin{eqnarray}
\dot{\eeta}=\textbf{J}\frac{\partial {\cal H}}{\partial
\eeta},\label{eq48}
\end{eqnarray}
\stepcounter{sub}
 where ${\cal H}(\eeta)$ is the extended
hamiltonian of Eq (\ref{eq9}) and $\textbf{J}$ is the symplectic
metric
\begin{eqnarray}
\textbf{J}=\left(%
\begin{array}{cc}
  j & 0\\
  0 & j \\
\end{array}%
\right), ~j=\left(%
\begin{array}{cc}
  0 & 1 \\
  -1 & 0 \\
\end{array}%
\right).\label{eq49}
\end{eqnarray}
\stepcounter{sub}
 An infinitesimal canonical transformation from
$\eeta$ to $\eeta+\delta\eeta$ is of the form
 \begin{equation}
\eeta+\delta\eeta=\eeta-\epsilon\frac{\partial G(\eeta)}{\partial
\eeta}~,\label{eq50}
\end{equation}
\stepcounter{sub}
 where $G$ is the generator of the transformation and
$\epsilon$
 indicates its infinitesimal character. The matrix of
the transformation is
\begin{equation}
M_{ij}=\delta_{ij}-\epsilon\frac{\partial^2 G}{\partial \eta_i
\eta_j} ~.\label{eq51}
\end{equation}
\stepcounter{sub}
 The condition for canonicity is
\begin{equation}
\textbf{M}\textbf{J}\textbf{M}^\dag=\textbf{J}+O(\epsilon^2).\label{eq52}
\end{equation}
\stepcounter{sub}
 This imposes the condition on $G$ to be either
linear in $\eta_j$ or quadratic and symmetric in $\eta_i$,
$\eta_j$ or both. For clarity, here after we confine our
discussion to a system of one degree of freedom, $N=1$. The most
general form of $G $ with the restriction just mentioned is
\begin{equation}
G(\eeta)=a_i\eta_i+\alpha_{ij}\eta_i\eta_j,~~~~i,j=1,2,3,4~\rm{correspond~
to~} q,~\pi_q;~p,~\pi_p,\label{eq53}
\end {equation}
\stepcounter{sub}
 where the four parameters $a_i$ initiate
translations and the 10 symmetric $\alpha_{ij}$ cause rotations,
boosts, squeezes, scale changes etc. The ten transformations
$\alpha_{ij}$ constitute a symplectic group $SP(4)$, and is
locally isomorphic to the $(3+2)$-dimensional Lorentz group. This
is the group that Kim and Noz \cite{KN} encounter in their study
of four dimensional phase space consisting of two oscillators
representation of $O(3+2)$, and proves to be a usefull
mathematical tool in quantum optics. To each of the fourteen
transformations of Eq (\ref{eq53}) there corresponds a unitary or
similarity transformation in the function space. That of Eq
(\ref{eq35}) it is unitary. An example of non unitary operators is
the following. To the canonical coordinate transformation
\begin{eqnarray}
q=Q+\frac{i\varepsilon}{2\hbar}\Pi_Q+\frac{1}{2}\Pi_P,~~~\pi_q=\Pi_Q\nonumber\\
p=P+\frac{i\hbar}{2\varepsilon}\Pi_P+\frac{1}{2}\Pi_Q,~~~\pi_p=\Pi_P,
\end{eqnarray}
\stepcounter{sub}
 there corresponds the complex similarity
operator
\begin{eqnarray}
S_\varepsilon=\exp{\left[(\frac{\varepsilon}{4}\frac{\partial^2}{\partial
q^2}+ \frac{\hbar^2}{4\varepsilon}\frac{\partial^2}{\partial
p^2})+\frac{i\hbar}{2}\frac{\partial^2}{\partial q \partial
p}\right]},
\end{eqnarray}
\stepcounter{sub}
 where $\varepsilon$ is a finite parameter of the transformation. Husimi's \cite{K. Husimi} all positive
distribution in terms of $\chi$ is
\begin{equation}
\chi_{Hus}(q,p,\varepsilon)=S_\varepsilon\chi(q,p).
\end{equation}
\stepcounter{sub}
 \section{\emph{Bloch's equation in phase space }}
 In this section we intend to illustrate some usage of the
 formalism developed so far.
In any discussion of statistical mechanics, the partition
function, $Z(\beta)$=tr$\hat{\Omega}$,
$\hat{\Omega}(q,\pi_q)=\exp(-\beta\hat{H})$, plays a pivotal role.
Its calculation, however, is often cumbersome. One practice is to
translate $\hat{\Omega}$ and the corresponding Bloch's
differential equation \cite{F. Bloch} into a phase space language
\cite{M. Hillery and R. F. O'Connell and M. O. Scully and E.
Wigner,R. F. O'Connell and L. Wang}, solve the equation for a
c-number $\Omega(q,p,\beta)$ and calculate
$Z(\beta)=\int\Omega(q,p,\beta)dqdp$. The ease of doing the job
depends on the choice of the c-number assigned to $\hat{\Omega}$.
Our suggestion is that of Eq (\ref{eq44}). Bloch's equation for
$\hat{\Omega}$ is
\begin{eqnarray}
 \frac{\partial \hat{\Omega}}{\partial
\beta}&=&-\hat{H}\hat{\Omega}=-\hat{\Omega}\hat{H}.\label{eq54}
\end{eqnarray}
\stepcounter{sub}
 We apply the rule of Eq (\ref{eq44}) to Eq
(\ref{eq54}). Noting that $\hat{H}(q,\pi_q)\rightarrow
H(q,p)=p^2/2m+V(q)$ and using the product rule of Eq (\ref{eq46})
gives
\begin{eqnarray}
-\frac{\partial\Omega(q,p;\beta)}{\partial
\beta}=\left\{H(q,p)-\frac{ip\hbar}{m}\frac{\partial}{\partial
q}-\frac{\hbar^2}{2m}\frac{\partial^2}{\partial
q^2}\right\}\Omega(q,p).\label{eq55}
\end{eqnarray}
\stepcounter{sub}
 This same result is obtained in \cite{N. L.
Balazs and B. K. Jennings}, however, by a totally different
approach and through much lengthier calculations using Moyal's
characteristic technique.
 Equation (\ref{eq54}) in Wigner's representation is obtained by replacing
 $\hat{\Omega}$ with $\Omega_W(q,p;\beta)$, $\hat{H}$ with $p^2/2m+V(q)$
 and using Eq (A17) with $\alpha=1/2$ to find the expression
 corresponding to $\hat{H}\hat{\Omega}$. In agreement with
\cite{M. Hillery and R. F. O'Connell and M. O. Scully and E.
 Wigner,R. F. O'Connell and L. Wang} one finds
\begin{eqnarray}
-\frac{\partial\Omega_W(q,p;\beta)}{\partial
\beta}=\left\{\frac{p^2}{2m}-\frac{\hbar^2}{8m}\frac{\partial^2}{\partial
q^2}+\sum_{n=0}^{n=\infty}\frac{(i\hbar/2)^n}{n!}
\frac{\partial^n}{\partial q^n}V(q)\frac{\partial^n}{\partial
p^n}\right\}\Omega_W(q,p;\beta).\label{eq56}
\end{eqnarray}
\stepcounter{sub}
 The contrast between the two Eqs (\ref{eq55})
and (\ref{eq56}) is striking. The former is a second order
differential equation in $q$ and the exact quantum effects in it
appear as $\hbar$ and $\hbar^2$ only, while the latter in addition
to $\frac{\partial^2}{\partial q^2}$ is an nth order differential
equations in $p$ and has all powers of $\hbar$ in it. In the
following we solve Eq (\ref{eq55}) and give the partition
functions for the simple harmonic and linear potentials.
\subsection{Simple harmonic potential}

For $H(q,p)=1/2(p^2/m+m\omega^2q^2)$ the solution is of the form
\begin{eqnarray}
\Omega=\exp{\left[-A(\beta)H(q,p)-iB(\beta)\frac{pq}{\hbar}-C(\beta)\right]}.\label{eq57}
\end{eqnarray}
\stepcounter{sub}
 Substituting this in Eq (\ref{eq55}) and letting
the coefficients of different powers of $q$ and $p$ vanish, gives
\setcounter{subeqn}{1}
\begin{eqnarray}
&&\frac{dA}{d\beta}=1-\hbar^2\omega^2A^2,\label{eq58}\\
\stepcounter{subeqn} && \frac{dB}{d\beta}=\hbar^2\omega^2A(1-B),
\label{eq59}\\
 \stepcounter{subeqn}
&&\frac{dC}{d\beta}=-\frac{1}{2}\hbar^2\omega^2A.\label{eq60}
\end{eqnarray}
\stepcounter{sub}
 The condition $\Omega(q,p,0)=1$ imposes the
boundary conditions $A(0)=B(0)=C(0)=0$. With these provisions one
finds \setcounter{subeqn}{1}
\begin{eqnarray}
&&A(\beta)=\frac{1}{\hbar\omega}\tanh \beta\hbar\omega,
\label{eq61}\\
\stepcounter{subeqn}
 && B(\beta)=\tanh
\beta\hbar\omega\tanh\frac{\beta\hbar\omega}{2},\label{eq62}\\
\stepcounter{subeqn}
&&C(\beta)=-\frac{1}{2}\ln\cosh\beta\hbar\omega.\label{eq63}
\end{eqnarray}
\stepcounter{sub}
The partition function is
\begin{eqnarray}
Z(\beta)=\frac{1}{2\pi\hbar}\int_{-\infty}^{+\infty}\Omega(q,p,\beta)dqdp=\left[
2\sinh\frac{\beta\hbar\omega}{2}\right]^{-1},\label{eq64}
\end{eqnarray}
\stepcounter{sub}
 The normalized density
function is $\chi(q,p,\beta)=\Omega(q,p,\beta)/2\pi\hbar
Z(\beta)$, with low and high temperature limits
\begin{eqnarray}
\chi&=&\frac{\sqrt{2}}{\pi\hbar}\exp[{-H(q,p)-\frac{ipq}{\hbar}}]~~~~~\beta\hbar\omega\gg1\nonumber\\
&=&\frac{\beta\omega}{2\pi}\exp{[-\beta
H(q,p)]}~~~~~~~~~\beta\hbar\omega\ll1,\label{eq65}
\end{eqnarray}
\stepcounter{sub}
 in agreement with the quantum and classical
limits, respectively.

\subsection{Linear potential}

The case is of interest for quark model \cite{J.Lin.Xu}, where a
sea of semi infinite matter creates a linear potential $V(q)=kq$,
$0\leq q<\infty$, and $k>0$. By the same procedure above one
obtains
\begin{eqnarray}
\Omega(q,p,\beta)=\exp{\left[-\beta H-\frac{i\beta^2 p\hbar
k}{2m}+\frac{\beta^3\hbar^2k^2}{6m}\right]},\label{eq66}
\end{eqnarray}
\stepcounter{sub}
\begin{eqnarray}
Z(\beta)=\sqrt{\frac{2\pi
m}{\beta^3k^2}}~\exp{\left[\frac{\beta^3\hbar^2k^2}{24m}\right]}.\label{eq67}
\end{eqnarray}
\stepcounter{sub}
\begin{eqnarray}
\chi(q,p~;\beta)&=&\sqrt{\frac{\beta^3k^2}{2\pi
m}}\exp{\left[-\beta H-\frac{i\beta^2p\hbar k}{2m}
+\frac{\beta^3\hbar^2k^2}{8m}\right]}.
\label{eq68}
\end{eqnarray}
\stepcounter{sub}
 The corresponding Wigner's function \cite{M.Durand et.al} can be
obtained by letting  $U_{1/2}$ operate on Eq (\ref{eq68}).

\section{Conclusion}

 We have developed a quantum mechanics in phase space by carrying
 the independent and symmetric roles of $q$ and $p$, so eminent in
 the hamiltonian formulation of the classical mechanics, to quantum
 domain. This is done through the extension of the phase space by
 introducing the momenta $\pi_q$ and $\pi_p$ conjugate to $q$ and
 $p$, respectively and the subsequent extensions of the lagrangians,
 hamiltonians, Poisson's brackets, etc. In its full generality,
 the theory describes the dynamics of the quantum ensembles. Its
 pure state case is reducible to the conventional quantum
 mechanics in $q$- or $p$-spaces, however, with a definite rule
 for  ordering of the factors of non commuting operators. The latter
 feature is a direct consequence of the independence of $q$ and
 $p$ that is maintained at all stages of the formalism. Simple
 rules for assigning an operator $\hat{F}(q,\pi_q)$ in $q$-space
 to a function $F(q,p)$ in phase space and vice versa are
 prescribed. Extended canonical transformations enable one to go
 from one extended phase space to another. Correspondingly the
 associated unitary or similarity transformations in the function
 space enable one to generate further state functions from a given
 one. This unifying feature of the theory makes the comparison of
 the various functions existing in the literature possible and
 transparent.

 To demonstrate the simplicity and the power of the formalism
 certain examples are worked out. Treatment of Bloch's equation,
 partition functions for simple harmonic and linear potentials, and
 the mathematical lemmas of the Appendix serve this end. Nasiri
 $\&$ Safari \cite{S. Nssiri and H. Safari} and Razavi \cite{M. Razavy} have found the presented
 formalism of considerable assistance in their study of
 dissipative quantum systems.

\begin{flushleft}
\textbf{Acknowledgments:} We thank M. R. H. Khajehpour for
fruitful discussions on all aspects of the paper and R. F.
O'Connell for valuable comments on our treatment of Bloch's
equation.
\end{flushleft}

\appendix
\section{} \label{ap1}
\begin{itemize}

\item \textbf{Evaluation of $\chi_\alpha=U_\alpha\chi$} :

 By three Fourier and
inverse Fourier transformations we convert
$\chi=\psi(q)\phi^*(p)\exp({-ipq/\hbar})$ into the following forms
$$
\chi=\int \phi(p')\phi^*(p'')\exp{\frac{iq(p'-p'')}{\hbar}}
\exp{\frac{ip\tau}{\hbar}}\exp{\frac{-ip''\tau}{\hbar}}dp'dp''d\tau.\eqno{(\rm{A}1)}
$$
By this provision we have moved both $q$ and $p$ variables to the
exponent. Next we expand $U_\alpha$ of Eq (\ref{eq36}) in power
series, operate by it on the $q$ and $p$ exponents and arrive at
$$
\sum_{k=0}^{k=\infty}\frac{(i\alpha\hbar)^k}{k!}\frac{\partial^k}{\partial{p}^k}
\frac{\partial^k}{\partial{q}^k}\exp{\frac{iq(p'-p'')}{\hbar}}\exp{\frac{ip\tau}{\hbar}}=
\exp{\frac{-i\alpha\tau(p'-p'')}{\hbar}}\exp{\frac{iq(p'-p'')}{\hbar}}\exp{\frac{ip\tau}{\hbar}}.\eqno{(\rm{A}2)}
$$
 Using Eqs (A1) and (A2) in the expression
$\chi_\alpha=U_\alpha\chi$, and inverting $\phi$'s back to
$\psi$'s gives Eq (\ref{eq37})
$$
\chi_\alpha(q,p,t)=U_{\alpha}\chi=(\frac{1}{2\pi\hbar})^N\int
\psi(q-\alpha\tau)\psi^*(q+(1-\alpha)\tau)e^{ip\tau/\hbar}d\tau.\eqno{(\rm{A}3)}
$$
As mentioned earlier, for $\alpha=1/2$ one recovers Wigner's
standard state functions.Q.E.D.

\item \textbf{ Evolution equation for $\chi_\alpha$ }

To prove Eq (\ref{eq38}), it is sufficient to evaluate ${\cal
H}_\alpha=U_\alpha {\cal H}U_\alpha^\dag$, where ${\cal H}$ is the
extended hamiltonian of Eq (\ref{eq12}). It is easy to show that
$Q=U_\alpha qU_\alpha^\dag=q-\alpha\pi_P$ and $P=U_\alpha p
~U_\alpha^\dag=p-\alpha\pi_q$, which is the essence of the
transformations of Eq (\ref{eq35}). We also note that
$$
U_\alpha q^n p^mU_\alpha^\dag=(q-\alpha\pi_p)^n(p-\alpha\pi_q)^m=
(p-\alpha\pi_q)^m(q-\alpha\pi_p)^n.\eqno{(\rm{A}4)}
$$
We leave it to the reader to verify Eq (A4) for
him/herself for some small $n$ and $m$. It is needless to say that
$[Q,P]=0$, because the transformation is unitary. With these
provisions one finds
$$
{\cal H}_\alpha=H\left[q-\alpha\pi_p,
p+(1-\alpha)\pi_q\right]-H\left[q-(1-\alpha)\pi_p,
p-\alpha\pi_q)\right].\eqno{(\rm{A}5)}
$$
 Expansion of the hamiltonian about $(q,p)$ gives
 $$
{\cal
H}_\alpha=-\frac{{\hbar}^2(1-2\alpha)}{2m}\frac{\partial^2}{\partial
q^2}-i\hbar\frac{p}{m}\frac{\partial}{\partial
q}+\sum_{n=0}\frac{(-\alpha)^n-(1-\alpha)^n}{n!}(-i\hbar)^n
\frac{\partial^{n}V}{\partial q^{n}}\frac{\partial^n}{\partial
p^n}.\eqno{(\rm{A}6)}
$$
The Wigner case is for $\alpha=1/2$
$$
{\cal H}_W=-i\hbar\frac{p}{m}\frac{\partial}{\partial
q}+\sum_{n=0}\frac{1}{(2n+1)!}(\frac{\hbar}{2i})^{2n+1}
\frac{\partial^{2n+1}V}{\partial
q^{2n+1}}\frac{\partial^{2n+1}}{\partial
p^{2n+1}}.\eqno{(\rm{A}7)}
$$Q.E.D.

\item \textbf{Ordering rule in $\alpha$-representation, proof of
Equation (\ref{eq41})}

With Eqs (\ref{eq39}) and (A3) we have
$$
<q^np^m>_{\alpha}=\int_{-\infty}^{+\infty} q^n
\psi(q-\alpha\tau)\psi^*(q+(1-\alpha)\tau)p^m
e^{ip\tau/\hbar}dqdpd\tau.\eqno{(\rm{A}8)}
$$
Writing $p^m$ as $(i\hbar)^m \frac{\partial^m}{\partial \tau^m}$,
integrating by parts $m$ times with respect to $\tau$ frees the
integrand from the $p^m$ factor. Then integration with respect to
$p$ gives $\delta(\tau)$. Thus
$$
<q^np^m>_{\alpha}=\int_{-\infty}^{+\infty}q^n(-i\hbar)^m
\frac{\partial^m}{\partial
\tau^m}\left[\psi(q-\alpha\tau)\psi^*(q+(1-\alpha)\tau)\right]\delta(\tau)
dqd\tau.\eqno{(\rm{A}9)}
$$
 Next we substitute $\frac{\partial}{\partial
\tau}$ by $\frac{\partial}{\partial q}$ with appropriate
adjustments and carry out integrations by parts over $q$ where
ever necessary to free $\psi^*$ and arrive at
$$
<q^np^m>_{\alpha}=\int_{-\infty}^{+\infty}\psi^*(q)\left[\sum_{r=0}^m\left(%
\begin{array}{c}
  m \\
  r \\
\end{array}%
\right) ((1-\alpha)\pi_q)^rq^n(\alpha\pi_q)^{m-r}\right]\psi(q) dq
.\eqno{(\rm{A}10)}
$$
The expression in the integrand is the desired ordering of Eq
(\ref{eq41}), corresponding to $q^np^m$ in
$\alpha$-representation. For $\alpha=1/2$ one recovers Weyl's
ordering
$$
q^np^m~\longrightarrow(\frac{1}{2})^m\sum_{r=0}^m\left(%
\begin{array}{c}
  m \\
  r \\
\end{array}%
\right) {\pi_q}^{r}q^n{\pi_q}^{m-r}.\eqno{(\rm{A}11)}
$$
The combination of $\alpha=0$ and $1$, corresponding to averaging
by $\chi+\chi^*$, is the symmetric ordering of Eq
(\ref{eq40}).Q.E.D.
\item  \textbf{The product rule, proof of Eqs (\ref{eq46}) and
(\ref{eq47})}

The phase space function corresponding to the product of two
operators $\hat{F}=\hat{A}\hat{B}$, by the definition of Eq
(\ref{eq44}), is
$$
F(q,p)=<q|\hat{A}\hat{B}|p>e^{-ipq/\hbar}=\int
<q|\hat{A}|p'><p'|q'><q'|\hat{B}|p>e^{-ipq/\hbar}dqdp $$
$$=\int A(q,p')B(q',p)\exp{\frac{-i(q'-q)(p'-p)}{\hbar}}dqdp,\eqno{(\rm{A}12)}
$$
where by Eq (\ref{eq44}), we have substituted
$<q|\hat{A}|p'>=A(q,p')\exp(ip'q/\hbar)$ and similarly for
$<q'|\hat{B}|p>$. With further change of variables $q'-q=q''$ and
$p'-p=p''$, we obtain
$$
F(q,p)=\int A(q,p''+p)B(q''+q,p)e^{-iq''p''/\hbar}dq''dp''=
\sum_{n=0}^{\infty}\frac{(-i\hbar)^n}{n!}\frac{\partial^nA(q,p)}{\partial
p^n}\frac{\partial^nB(q,p)}{\partial q^n},\eqno{(\rm{A}13)}
$$
where we have Taylor-expanded $A(q,p+p'')$ and $B(q+q'',p)$ about
$(q,p)$ and carried out the required integration by parts.

 To deduce Eq (\ref{eq47}), we first Fourier-transform $A(q,p)$
to $a(p',q')$ and $B(q,p)$ to $b(p'',q'')$ in Eq (A13) and carry
out the necessary differentiations:
$$
F(q,p)=\sum_{n=0}^{\infty}\frac{(-i\hbar)^n}{n!}\frac{\partial^n}{\partial
p^n}\int
a(p',q')\exp{\frac{-ip'q+iq'p}{\hbar}}dp'dq'\frac{\partial^n}{\partial
q^n}\int b(p'',q'')\exp{\frac{-ip''q+iq''p}{\hbar}}dq''dp''$$ $$=
\int
a(p',q')\exp{\frac{-ip'q+iq'p}{\hbar}}~b(p'',q'')\exp{\frac{-ip''q+iq''p}{\hbar}}~
e^{-iq'p''/\hbar}dq'dp'dq''dp''.\eqno{(\rm{A}14)}
$$
Next we operate on Eq (A14) by a Taylor-expanded form of
$U_\alpha$ as in Eq (A2) and perform the required
differentiations:
$$
F_\alpha(q,p)=U_\alpha F(q,p)=\int e^{i\alpha
q'p'/\hbar}\exp{\frac{-ip'q+iq'p-(1-\alpha)q'p''+\alpha
q''p'}{\hbar}} a(q',p')$$
$$\times e^{i\alpha
q''p''/\hbar}\exp{\frac{-ip''q+iq''p}{\hbar}}b(p'',q'')
dq'dp'dq''dp''.\eqno{(\rm{A}15)}
$$
The exponentials preceding $a(p',q')$ can be written as
$$
 e^{i\hbar\alpha \partial^2/\partial p \partial
q}\exp{\frac{-ip'(q+\alpha
q'')+iq'(p-(1-\alpha)p'')}{\hbar}}\nonumber,
$$
 where
the first factor is simply $U_\alpha(q,p)$ independent of the
integration variables $(q',~p',~q'',~p'')$. With this provision
integrations over $q'$ and $p'$ can now be carried out and
$a(p',q')$ inverse-Fourier transformed. One finds
$$
F_\alpha(q,p)=\int\left\{U_\alpha A[ q+\alpha
q'',p-(1-\alpha)p'']\right\}e^{i\alpha
q''p''/\hbar}e^{\frac{-ip''q+iq''p}{\hbar}}b(p'',q'')dq''dp''.\eqno{(\rm{A}16)}
$$
We again apply the same trick. To the left of the right most
exponential we replace, every where, $q''$ by $(-i\hbar
\partial/\partial p)$ and $p''$ by $(i\hbar
\partial/\partial q)$, perform the inverse Fourier transform of
$b(p'',q'')$ and find
$$
F_\alpha(q,p)=U_\alpha F(q,p)=A_\alpha\left[ q+i\hbar\alpha
\frac{\partial}{\partial
p},p-i\hbar\alpha(1-\alpha)\frac{\partial}{\partial
q}\right]B_\alpha(q,p)$$$$=\sum_{n=0}^{\infty}
\frac{(i\hbar)^n}{n!}\left[\alpha\frac{\partial }{\partial
q_A}\frac{\partial}{\partial p_B}-(1-\alpha)\frac{\partial
}{\partial p_A}\frac{\partial}{\partial
q_B}\right]^nA_\alpha(q,p)B_\alpha(q,p).\eqno{(\rm{A}17)}
$$

 where $\partial/\partial p_A$ indicates a differentiation on
$A(q,p)$ only, similarly the other differential operators. Q.E.D.
\end{itemize}

\end{document}